# Product Line Requirements Matching and Deriving: the RED-PL Guidance Approach

Olfa Djebbi[12], Camille Salinesi[1], Daniel Diaz[1]
[1] CRI, Université Paris 1 – Sorbonne, 90, rue de Tolbiac, 75013 Paris, France
[2] Satgo Instruments, 125 avenue Louis Roche, 92230 Gennevilliers, France
olfa.djebbi@malix.univ-paris1.fr, {Camille.Salinesi, Daniel.Diaz}@univ-paris1.fr,
odjebbi@stago.fr

**Abstract**

*Product lines (PL) modeling have proven to be an effective approach to reuse in software development. Several variability approaches were developed to plan requirements reuse, but only little of them actually address the issue of deriving product requirements.*

*This paper presents a method, RED-PL that intends to support requirements derivation. The originality of the proposed approach is that (i) it is user-oriented, (ii) it guides product requirements elicitation and derivation as a decision making activity, and (iii) it provides systematic and interactive guidance assisting analysts in taking decisions about requirements.*

*The RED-PL methodological process was validated in an industrial setting by considering the requirement engineering phase of a product line of blood analyzers.*

## 1. Introduction

Product Line Engineering (PLE) has proven to be a viable development paradigm that allows companies to realize order-of-magnitude improvements in time to market, cost, productivity, quality and flexibility. However, maintaining PL models is not enough to take benefit of these advantages. It is also crucial to maintain a marketplace and build products that will be sold. Therefore, a special attention should be paid to customers' needs and requirements while developing products, and while optimizing costs and other constraints.

Requirements Engineering (RE) processes have two main goals when put in the context of PLE: (i) to define and manage requirements within the product line and (ii) to coordinate requirements for individual products. The latter goal should be achieved in a very specific way as, contrary to the case of a new development, there are two kinds of requirements to be considered: those of the customers, and those that the product line is able to satisfy.

Some recommendations can be found to guide RE processes in the context of PLE [1] [2] [3] [4] [5]. In these approaches, the first of the aforementioned PLE goal is achieved by a requirements variability modeling approach. The second goal is usually achieved by a requirements selection approach. Requirements selection consists in building the collection of requirements for the product to build consistently with the requirements identified in the PL requirements variability model.

Several serious limitations of this way of working have been shown by [6] and [7]. These limits concern in particular the lack of representation of customers' requirements, and the RE process itself.

Selecting requirements among pre-defined product line requirements models influences stakeholders and skews their choices. Experience with this approach in other domains such as COTS selection or ERP implementation shows that stakeholders naturally establish links between their problem and the pre-defined solutions, adopt features with marginal value, and naturally forget about important requirements that are not present in the PL requirements model [8] [9]. As a result, the focus is on model elements that implement the solution rather than on the expression of actual needs. While this approach supports reuse, it generates products that finally lack of attractiveness, or even worse usefulness. Each important requirement missing leads to unsatisfied final users and customers.

Besides, while the RE process should foster creative thinking within the system requirements, selecting among predefined requirements restricts considerably creativity and search for innovative ways to deal with problems, and hence reduces the added value of the new products to be developed.

Moreover, analysts are most often on their own to elicit the requirements for new products. As shown in previous publications [10] [11], existing approaches and tools provide little guidance (notation, process, rules, impact analysis) to assist them in eliciting consistent product requirements. They are neither guided in adding new requirements to the PL requirements model to support more complex evolutions of the PL requirements model [12].

On the other hand, an approach in which stakeholders would come up with completely new requirements, specifying these independently from the PL requirements model would be difficult to handle and can become very inefficient. Indeed, retrieving correspondences between customers' requirements and PL requirements can be time taking, error prone, and implies to face difficult issues such as inconsistent levels of abstraction, inconsistency in the way similar requirements are expressed, and the need for large amount of details to decide whether customers' requirements are satisfied. We strongly believe that a systematic guidance is needed to facilitate this activity and, most importantly to check the consistency of product requirements with PL and customers' requirements models.

More precisely, we believe that a "good"[1] product requirements derivation approach should satisfy the following characteristics:

- *Requirements oriented*: customers should be able to express their real needs with as little external influence as possible; the product built should satisfy these requirements.
- *Product line based*: the developed product should take advantage of the PL platform and reuse elaborated requirements so as to be traced and validated.
- *Unified into the whole PL development cycle*: the approach should provide means to ensure traceability with the remaining development phases for both the product line and individual products.
- *Provide interactive guidance*: the derivation approach should integrate guidance assisting analysts in taking decisions about product requirements. This can take various forms such as impact analysis tools, wizards, informal guidelines, etc. One important aspect is to get an adequate kind of guidance for each situation in which guidance is needed.
- *Supported by a CASE tool* that is integrated into existing toolkits: appropriate tool support is mandatory to automate methodological processes, and hence their large adoption by developers' community.
- *Scalable*: the method should allow modeling real-scale systems.

---
[1] at least in the sense that it would face the aforementioned shortcomings of existing approaches

This paper presents ongoing research towards the development of a requirements derivation approach meeting these objectives. Our research strategy is experience based. It consisted in undertaking innovation/validation cycles in a practical study within an industrial company managing a product line of blood analysis automatons [13]. We proceeded by gradually introducing basic PL management principles in the RE phases of product creation projects and validating them by studying obtained results and consulting domain experts. Based on this experience, we developed a method, named RED-PL (Requirements Elicitation & Derivation for Product Lines), that guides the elicitation of product requirements by derivation from the PL requirements specification. RED-PL is based on already existing PL requirements notations. Its originality is that (i) it is user-oriented, and (ii) it guides product requirements elicitation and derivation as a decision making activity.

RED-PL makes it possible to users to express their needs using classic RE techniques. Mechanisms are proposed to convert these needs and match them with the PL requirements specification. Negotiation and arbitration are also supported in RED-PL to elicit optimal product requirements while maximizing reuse. This paper focuses the derivation part of the RED-PL approach. It provides guidelines for each step and hints for its implementation.

The remaining of the paper is structured as followed. Section 2 outlines derivation process related works, presents the outline of the proposed RED-PL approach and explains how it meets requirements derivation challenges cited above through its processes and guidance. Section 3 is dedicated to the technical implementation of RED-PL. And finally, conclusions and discussions about the validity of the approach and the future work are reported in section 4.

## 2. The RED-PL approach

### 2.1. Related works

Several methods guiding the construction of PL assets are available in literature [14] [15] [16]. Product derivation methodologies are on the contrary rather scarce [2] [17] [18]. Besides, although derivation affects the whole product line artifacts, from requirements to code, the derivation issues are mainly addressed in terms of design and implementation [2] [4]. Approaches that tackle the requirements level [5] [19] [20] [21] [22] [23] mostly deal with the creation of the right requirements assets for the PL and dependencies among them to develop the right products. Understanding the derivation process at the requirements level has received little attention.

In existing derivation approaches, the derivation of the product architecture, code or test artifacts from PL specifications is performed using the following techniques:
- *Model transformation*: static and dynamic models are instantiated for products from the PL models, using a model transformation language [2] [24] [25].
- *Design patterns*: are for instance used in Jezequel's method. This method consists in using the 'Abstract Factory' design pattern to create products [26].
- *Variability control*: generative approaches such as Generative Programming [17] guide automatic derivation by code generation. Selecting desired product features is sufficient to allow assembling correspondent PL elementary reusable components and generate the application code. Other approaches introduce aspect programming techniques to assemble components by waving features [27] [28].

For most of these derivation methods the input is a collection of PL requirements selected from the PL requirements model. However, industrials need more than just selecting assets from requirements. It is also necessary to be able to constrain requirements selection by pre-selected asset assemblies. For instance, a production plan that describes which core assets should be used to develop products must be systematically considered [29]. To achieve this, Hunt considers software components and studies the optimal organization to guide their identification and selection [30]. [31] discusses automated component selection using artificial intelligence techniques. [1] and [3] provide a framework and a generic process to guide software derivation, which is organized in iterative phases that determine the final configuration of the derived product. The input of the derivation process is a subset of the requirements originating from customers, legislation, hardware and product family organization. Unfortunately, the approach provides no detail on how these requirements should be aggregated. Another derivation framework is provided by [2]. In this framework, product requirements derivation is achieved through a decision process. Again, details are missing about the process to make it really systematic.

One important aspect of product derivation shown by this review is that determining the requirements for a particular product calls for (i) considering some sort of description of the customers' needs separately from the PL requirements (which are only the requirements that the product line is able to satisfy), (ii) considering the constraints imposed by the developers in terms of core assets to be used in the product.

While most of the approaches focus on handling technical derivation, we are interested in requirements derivation process that conciliates customers' needs, technical constraints, and guides decision making in a systematic way.

## 2.2. Outline of the RED-PL approach

As depicted in Figure 1, RED-PL supports three goals:
- *Elicit stakeholders' requirements*. By stakeholder we mean customers and users, as well as actors from the developing organisation: strategy makers, marketing, engineers, developers, etc.
- *Match stakeholders' requirements with PL requirements*. This establishes the collection of requirements that are both covered by the PL and that satisfy stakeholders' needs. To this collection of requirements usually corresponds a set of products that are consistent with the PL requirements model; a choice must thus be made. Matching can also lead to capitalizing on new requirements by introducing them in the PL requirements model.
- *Derive the optimal collection of product requirements*. This is achieved by taking into account different kinds of constraints that were not covered by stakeholders requirements such as cost, development time, risk, etc. Non functional requirements such as flexibility or maintainability can be used as decision criteria too.

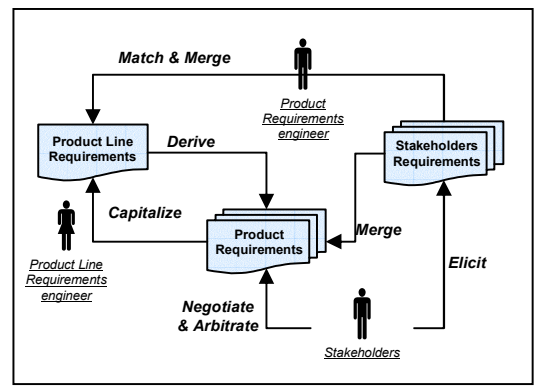

Figure 1. **Overview of the RED-PL approach**

Several intertwined processes are guided by the RED-PL approach: requirements elicitation, matching, merging, deriving, capitalization, negotiation and arbitrating. These are achieved as follows.

In the RED-PL approach, analysts assist stakeholders in eliciting their requirements regarding the new product using classical requirements elicitation techniques such as structured interviews, Use Case analysis or goal modeling. Applying well established methods allows to focus on the real stakeholders needs and to ensure correctness, completeness, and consistency.

### 2.3. Matching technique

Requirements are then interpreted and matched to the PL requirements. Requirements' matching uses similarity analysis techniques and calls for reformulation when conceptual mismatch issues are met. Two kinds of similarity analysis techniques can be used: surface level and deep level. Surface level techniques are based on lexical similarity: two requirements are considered similar when they use the same term. Deep level technique uses a structural and a semantic proximity. This allows to identify similar requirements, even though they are not expressed using the same terms or using the same linguistic structures. These techniques need more sophisticated tools such a dictionaries and linguistic parsers. Our similarity analysis approach also uses refinement, as suggested by goal modeling, to progressively improve the quality of the matching and to focus on requirements that are considered more important [32].

Our approach exploits the 30 generic similarity metrics developed by [8] and adapted to Dice, Jaccard and Cosine's ratios. As shown below, similarity can be automatically computed by applying a weighted ratio between a number of similarities found between two requirements and the number of elements that define these requirements.

$$S_D^m(A,B) = \frac{\sum_A MAX_B[SIM(Termes_A, Termes_B)] + \sum_B MAX_A[SIM(Termes_A, Termes_B)]}{|\{Termes_A\}| + |\{Termes_B\}|}$$

(Formula 1) *Adapted Dice ratio*

$$S_J^m(A,B) = \frac{\frac{1}{2}\sum_A MAX_B[SIM(Termes_A, Termes_B)] + \frac{1}{2}\sum_B MAX_A[SIM(Termes_A, Termes_B)]}{|\{Termes_A\}| + |\{Termes_B\}| - \left(\frac{1}{2}\sum_A MAX_B[SIM(Termes_A, Termes_B)] + \frac{1}{2}\sum_B MAX_A[SIM(Termes_A, Termes_B)]\right)}$$

(Formula 2) *Adapted Jaccard ratio*

$$S_C^m(A,B) = \frac{\frac{1}{2}\sum_A MAX_B[SIM(Termes_A, Termes_B)] + \frac{1}{2}\sum_B MAX_A[SIM(Termes_A, Termes_B)]}{\sqrt{|\{Termes_A\}| \times |\{Termes_B\}|}}$$

(Formula 3) *Adapted Cosine's ratio*

These ratios use a SIM function that computes the similarity between simple items as follows:
SIM(A,B) = 1     if A and B are identical,
SIM(A,B) = 1-a   if A and B are homonyms,
SIM(A,B) = 1-b   if A is an hyponym of B,
SIM(A,B) = b     if b is a hyperonym of A,
SIM(A,B) = 0     otherwise,
with a and b between 0 and 1 excluded.

The matching process is an iterative process that results in a merged collection of requirements that shall be implemented in the product. Requirements merging is achieved by (i) fetching and mapping original stakeholders' requirements into the PL requirements model, (ii) revising PL requirements with new ones, (iii) negotiating on whether to include requirements in the derived product specifications, (iv) reformulate the stakeholders' requirements model if a conceptual mismatch issue is met, and (v) re-iterate until a sufficient level of detail is found.

This activity allows *refining progressively the final product requirements* while relating to the PL capabilities and assessing reuse, as well as updating the PL assets. To the product requirements thus obtained corresponds a subset of all suitable products. The arbitration process is applied to derive the optimal set of product requirements, with respect to environmental, user's and company's constraints.

We observed in current industrial practices, that to achieve this, guidance is needed to assist stakeholders and analysts in taking decisions while achieving these processes. Several questions are typically raised throughout the derivation cycle: "How can I compose the cheaper product configuration? If I choose this requirement to be part of the new product, is it restrictive for remaining choices? If I add a new requirement and its corresponding dependency relationships in the PL model, can I be assured that my model is kept consistent? I have some requirements that I must include in the product configuration, how can I deduce the remaining requirements in such a way that I make a valid product model arranging my priorities and constraints?" Answering these questions is time consumed, tedious and risk-prone due to their combinatorial nature.

We propose to deal with this need for guidance by a wizard as described in the next section.

## 2.4. Derivation wizard

Derivation is carried out through interactive queries. The objective is twofold: (i) to allow stakeholders make decisions with a good view on their impacts, and (ii) to control the validity of stakeholders' decisions. Two kinds of queries can be distinguished: those that relate to model validity, and those that relate to requirements selection.

**Model validity queries:**

Queries that concern the validity of model can apply to product requirements models, and to PL requirements models. Our choice in RED-PL was to support feature modeling. Therefore, the queries dealt with by RED-PL apply to Feature models. Similar queries would emerge if another kind of modeling language was used.

In so far as PL models are concerned, guidance is needed to check that:
- the model is an acyclic oriented graph
- the model doesn't contain an isolated feature
- within the model, a feature can be of three types: either a mandatory feature, or an optional one, or belonging to a group of features
- within a model, a group of features has one and only one cardinality
- a model does not contain contradictory dependency relationships or feature types, e.g. 'mutex' and 'requires' relations between the same features, an optional feature 'required' by a mandatory one, etc.
- at least one valid configuration can be derived from the model. Details concerning this query are precisely reported below.

A product configuration is an extract of the PL model, so it must verify all previous validity conditions. Besides, a valid configuration is a product model that verifies the additional following conditions:
- the model contains all PL mandatory features
- the variation points of the model are well solved, i.e. options and cardinalities are rightly considered
- the dependency relationships of the model are respected

**Requirements selection queries:**

Beyond validity queries, requirements selection queries are the core of the derivation guidance. Their resolution should vastly help analysts in conducting the requirements derivation process. In front of his models, an analyst may ask:
- What are the possible configurations that can be composed from a given PL model?
- What are all the possible configurations that include a pre-selected set of requirements?
- What are the configurations that do not contain a given requirement to be excluded?
- What are the requirements that respect a given criterion? (e.g. what are requirements with 'implementation cost'<c?)
- What is the optimal requirements configuration with respect to a criterion such as the minimal total cost configuration, minimal number of features, etc?
- Given an initial requirements selection, what are the choices that still need to be made?
- Is a given requirement consistent with all the requirements already adopted?

This list is not exhaustive. Other user-defined queries can be defined by personalizing requirements attributes and their values, or by combining requests, as for example in:
- what are all alternative requirements with 'cost'<c?
- what are all possible configurations with respect to a pre-selection of some requirements following some criteria?
- having excluded some requirements, what is the cheaper configuration among reminder ones?

One important challenge to handle these questions is obviously to efficiently check both the constraints underlying the model and the constraints expressed by users for the product under development.

## 3. Tool Implementation

So far, our techniques have only been applied manually, which implied both to follow the RED-PL methodological process without assistance, and to implement the techniques and achieved the required calculations on the fly. We intend to guide RED-PL using an interactive tool that would have the following key features:

(1) A classical graphical editor that lets draw, load and save PL requirements models, models of early stakeholders' requirements, and actual requirements for the individual products. Such an editor should be linked to a shared repository with concurrent access and version control facilities.

(2) An efficient model validity checking feature. Incremental maintenance should be possible as long as models are modified.

(3) An intuitive way to express wishes about the product to derive: the user should be able to select/exclude some features in order to build a product configuration (but we can imagine more complex expressions).

(4) An efficient computation of a first complete solution w.r.t. the selected/excluded features, so as to provide a general idea of the product that is built.

(5) A efficient next solution computation that offers an alternative to the previous solution. Iterating over this function allows to review the various solutions one by one.

(6) Interactive guidance of product construction. This should consist helping users completing step by step a partial solution starting from stakeholders' requirements. Each time a new stakeholder' requirement emerges, a matching is made with PL requirements model as defined earlier. Once the user has selected/excluded a feature, all deductible consequences should be automatically shown under the form of decisions that remain to be made. If no matching can be made, then the suggestion shall be to refine or revise the stockholder's requirement.

In the above description, each computation should be as efficient as possible since the tool interacts with the user (and the user does not like to wait for a too long time). The main challenge is then to efficiently handle both the constraints associated to the underline model drawn as in (1) and the constraints expressed by the user for the final product as in (3).

If we were only interested in finding a complete solution as for (4), any simple constraint solver could be used. This problem can be solved by: pure boolean methods, Operational Research techniques for Integer Linear Programming (ILP), or Consistency techniques designed for Constraint Satisfaction Problems (CSP).

We experimented the ILP approach [33] by associating a 0-1 variable to each feature whose value indicated if the corresponding feature should be present in the final product. The model was then translated as a set of constraints over those variables as shown in the following table (which also details propagation rules):

| Dependency | Translation |
|---|---|
| 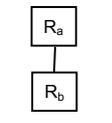 (composition) | If a requirement is selected then all mandatory requirements composing it must be selected <br><br> Constraint: $R_a = R_b$ |
| 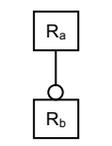 (option) | If a requirement is selected then its sub-requirements may be selected <br> $R_a = 0 \Rightarrow R_b = 0$ <br> $R_b = 1 \Rightarrow R_a = 1$ <br><br> Constraint: $R_b \leq R_a$ |
| 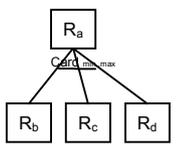 (alternatives) | If a requirement is selected then alternative sub-requirements must be selected respecting the specified cardinality <br> $R_a = 1 \Rightarrow$ <br> $R_b + R_c + R_d \leq Card_{max}$ and <br> $R_b + R_c + R_d \geq Card_{min}$ <br> $R_a = 0 \Rightarrow R_{b..d} = 0$ <br> $R_{b..d} = 1 \Rightarrow R_a = 1$ <br><br> Constraints: $R_{b..d} \leq R_a$ <br> $R_a * Card_{min} \leq R_b + .. + R_d \leq Card_{max}$ |
| 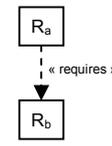 (requires) | If a requirement is selected then a required requirement must be selected <br> $R_a = 1 \Rightarrow R_b = 1$ <br> $R_b = 0 \Rightarrow R_a = 0$ <br><br> Constraint: $R_a \leq R_b$ |
| 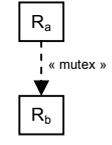 (mutex) | If a requirement is selected then a requirement mutually exclusive with it must not be selected <br> $R_a = 1 \Rightarrow R_b = 0$ <br> $R_b = 1 \Rightarrow R_a = 0$ <br><br> Constraint: $R_a + R_b \leq 1$ |

This ILP approach was applied to a real case developed in the STAGO company. The experiment showed that this approach could be used without initial constraints or with a simple pre-selection of features, but not in the case where complex requirements had to be expressed. It also showed scalability problems.

Recently [34] performed a performance comparison between two boolean methods (one based on BDD - Binary Decision Diagrams- and another using a SAT – boolean SATisfiability- method) and one CSP method. These methods were tested on a set of randomly generated benchmarks. The paper only focuses on finding one complete solution as needed for and counting the total number of solutions. At a first reading one could conclude that the BDD approach outperforms the two other challengers. But the authors themselves conclude that there is not an optimum representation for all the possible operations that can be performed on feature models. Obviously, the well-known NP-completeness of the satisfiability of boolean formulas shows that we are tackling a difficult problem here in the general case. However, our experience on solving boolean constraints showed that CSP techniques can solve many problems that cannot be handled with BDD [35]. We also know that results obtained on random problems can be very far from what we obtain in practice on real-life applications.

While these approaches satisfy the issue raised by (4), they cannot handle more critical issues such as those raised by (5) and (6). Indeed, these requirements imply the use of a more complex solver since we are also interested in incrementality dealing with a partial solution and in finding several solutions. Besides, we want a tool that is enough flexible to be able to deal with new needs such as the introduction of new types of requirement dependencies, new kinds of constrains, or richer way to express stakeholders' requirements (e.g. "we want at most 3 occurrences of feature X in a final product"). It appears that Constraint Programming is the most adequate paradigm to fulfill all these requirements.

Constraint programming is a powerful paradigm for solving combinatorial problems arising in many domains, such as scheduling, planning, vehicle routing, configuration, networks or bioinformatics. The idea of constraint programming is to solve problems by stating constraints and finding a solution satisfying all the constraints. A constraint is simply a logical relation between several unknowns, these unknowns being variables that should take values in some specific domain of interest. A constraint thus restricts the degrees of freedom (possible values) the unknowns can take; it represents some partial information relating the objects of interest. The execution of a program mainly adds the constraints (incrementally) and asks the built-in solver to find a solution (an assignment of variables that satisfies the constraints). Constraint Programming really appeared in the context of Logic Programming in the 80s to give rise to Constraint Logic Programming (CLP) [36]. Constraints were smoothly integrated into Logic Programming since the unification (an equation over trees) is a particular case of constraint (equality) on a given domain (syntactic trees). The resulting CLP(X) framework is parameterized by a constraint system X. Classical systems include Reals, Intervals, Rationals, Booleans, Finite Domains for arithmetics as well as Rational Trees, Lists and Sets [37]. In fact, X can be any system respecting some properties and for which there exists an efficient solving algorithm. Several CLP systems were designed like CHIP [38] and GNU-Prolog [39] for Finite Domains, clp(R) for Reals [40], Prolog-III for Reals, Trees and Lists [41].

Constraint Programming has grown and is no longer limited to Logic Programming: several libraries implementing constraint solving are available for languages like C, C++, or Java. Constraint Programming has been identified by the ACM (Association for Computing Machinery) as one of the strategic directions in computer research.

Constraint Programming over Finite Domains is clearly the most adequate way to implement our list of requirements. A Finite Domain variable is a variable whose initial domain is a finite set of integers. The Finite Domain constraint system offers to the user a wide variety of constraints. For instance the solver we have developed in GNU Prolog [39] offers:

• arithmetic constraints (both linear and non-linear). e.g. $X+Y \leq Z$ or $X*Y \neq Z$.

• symbolic constraints. e.g. atmost (2,[X,Y,Z,T],10) states that at most 2 variables among X,Y,Z,T can take the value 10.

• reified constraints: making it possible to reason on the issue of a constraint. e.g. with $X<Y \Rightarrow K=8$ as soon as the solver discovers $X<Y$ it enforces $K=8$ (conversely as soon as it detects $K \neq 8$ it enforces $X \geq Y$).

These constraints are mainly solved by consistency techniques issued from CSP [38] [42] but also with techniques borrowed from Operations Research. In addition, GNU Prolog also offers various enumeration heuristics and optimization facilities. Probably more than 90% of all industrial constraint applications use Finite Domains. Obviously a boolean variable is a special case of Finite Domain and we have shown how to efficiently solve boolean problems with a Finite Domain solver [35]. It is worth noting that dealing with Finite Domain makes it possible to enrich both the model (1) and the wishes (3) if we discover new needs. Here are some examples:

- We could imagine more complex constraints between 2 (or more) features. For instance: "feature X is mutually exclusive with feature Y if some computed information on X is greater than 10".
- We could allow the user to specify a wish like "I want at most 2 occurrences of feature X" in the final product.
- If each feature has a weight we could ask for a product whose total weight is less than 5 Kgs.
- We could add a cost (and/or a benefit) to each feature and ask for a product minimizing some objective function over those costs (benefits).

Constraint Programming not only brings us the needed efficiency for the resolution but also the essential flexibility for such a tool (whose requirements could evolve when attacking real-life problems).

## 4. Conclusion

A major addition to existing reuse approaches since the 90s are product line paradigm that has been the long standing notion to solve the cost, quality and time-to-market issues associated with development of related applications.

Over the past few years, domain engineering has received substantial attention from the software engineering community. Most of the researches, however, fail to provide detailed derivation processes namely for deriving requirements, which has been restricted to the selection of a requirements subset from the PL assets.

The idea behind the proposed approach in this paper is that:

(i) the user, the main stakeholder to whom the final product is intended, should be involved in specifying product requirements, in a way that efforts expended in constructing the reusable requirements in domain engineering are outweighed by the benefits in deriving the right individual products that satisfy their mission.

(ii) the analysts who conduct the derivation process should dispose of formal processes and automatic means to be able to take efficient decisions about product requirements to build.

The proposed RED-PL approach makes it possible to users to express their needs using classic RE techniques. Mechanisms are proposed to convert these needs and match them with the PL requirements specification using similarity analysis techniques. This establishes the collection of requirements that are both covered by the PL and that satisfy stakeholders' needs. Matching can also lead to capitalizing on new requirements by introducing them in the PL requirements model.

To the obtained collection of requirements usually corresponds a set of products that are consistent with the PL requirements model. Negotiation and arbitration are thus also supported in RED-PL in order to derive a consistent and optimal product requirements taking into account different kinds of constraints that were not covered by stakeholders requirements.

As observed in current industrial practices, guidance is needed to assist stakeholders and analysts in taking decisions while achieving these processes. So RED-PL deal also with this need for guidance by a derivation wizard described as a set of interactive queries allowing stakeholders making decisions with a good view on their impacts, and controlling the validity of these decisions.

Besides, we intend to support RED-PL systematic guidance using an interactive tool. It should allow following the RED-PL methodological processes and enable assistance by implementing techniques to achieve query calculations on the fly. We demonstrate that Constraint Programming is the most adequate paradigm to fulfill this.

Our research strategy is experience based. It consisted in undertaking innovation/validation cycles in a practical study within an industrial company managing a product line of blood analysis automatons [13]. We proceeded by gradually introducing basic PL management principles in the RE phases of product creation projects and validating them by studying obtained results and consulting domain experts.

Further research will focus on the refinement of the RED-PL approach processes. We aim also at implementing the tool supporting it and that can be interfaced with existing modeling tools.